%% file: EM_velocity.tex
\documentclass[11pt]{article}
\usepackage{amssymb,amsmath} 
\usepackage{graphicx}
\usepackage[colorlinks=true, linkcolor=blue, citecolor=blue, urlcolor=blue]{hyperref}

\input B3macros.tex

\def\bib#1{\bibitem[#1]{#1}}

\begin{document}

\title{Electromagnetic inertia, reactive energy,\\
and energy flow velocity}

\author{Gerald Kaiser\\
\href{http://wavelets.com}{Center for Signals and Waves}\\ Austin, TX\\
kaiser@wavelets.com
}

\maketitle


%

\maketitle

\begin{abstract}\noindent 
In a recent paper titled \href{http://arxiv.org/abs/1102.0238}{Coherent electromagnetic wavelets and their twisting null congruences}, I defined the local \sl  inertia density \rm $\5I\xt$,  \sl reactive energy density \rm $\5R\xt$, and  \sl energy flow velocity \rm $\3v\xt$ of an electromagnetic field. 
These are the field equivalents of the \sl mass, rest energy, and velocity \rm of a relativistic particle. Thus $\5R=\5Ic^2$ is Lorentz-invariant and $|\3v|\le c$, with equality if and only if $\5R=0$. The exceptional fields with $|\3v|=c$ were called \sl coherent \rm because their energy moves in complete harmony with the field, leaving no inertia or reactive energy behind. Generic electromagnetic fields become coherent only in the far zone. Elsewhere, their energy flows at speeds $v\xt<c$, a statement that is surprising even to some experts. The purpose of this paper is to confirm and clarify this statement by studying the local energy flow in several common systems: a time-harmonic electric dipole field, a time-dependent electric dipole field, and a standing plane wave. For these fields, the energy current (Poynting vector) is too weak to carry all of the energy, thus leaving reactive energy  in its wake. For the time-dependent dipole field, we find that the energy can flow both \sl transversally \rm and \sl inwards, \rm back to the source. Neither of these phenomena show up in the usual computation of the \sl energy transport velocity \rm which considers only averages over one period in the time-harmonic case.

\end{abstract}

\VE

\tableofcontents

\section{Introduction}\label{S:inertia}

Given an electromagnetic field $(\3E,\3B)$ in vacuum, its \sl energy density \rm $\5U$ and \sl energy current  (Poynting's vector) \rm $\3S$ are defined in Heaviside-Lorentz units $(\e_0=\m_0=1)$ by
\begin{align}\lab{ES}
\5U\0x=\frac12(\3E\0x^2+\3B\0x^2)\ \ \hbox{and}\ \ \3S\0x=\3E\0x\times\3B\0x,
\end{align}
where $x=\xt\in\rr4$ are the spacetime coordinates. By Maxwell's equations, the four-vector field $(\3S,\5U)$ satisfies Poynting's conservation law
\begin{align}\lab{PoynThm}
\pl_t\,\5U+c\div\3S=-\3E\cdot\3J,
\end{align}
where $c$ is the speed of light and $\3J$ is the current density. In the absence of a current, energy is conserved. Due to the vector identity
\begin{align*}
(\3E\times\3B)^2=\3E^2\3B^2-(\3E\cdot\3B)^2,
\end{align*}
we have
\begin{align}\lab{pos}
\5U^2-\3S^2=\frac14(\3E^2-\3B^2)^2+(\3E\cdot\3B)^2\ge 0.
\end{align}
At every event $x$, $(\3S,\5U)$ is either \sl future timelike \rm ($\5U\0x>|\3S\0x|$) or \sl future lightlike \rm  ($\5U\0x=|\3S\0x|$).

The inequality \eq{pos} does not depend on Maxwell's equations. It is satisfied by \sl any two vectors \rm $(\3E, \3B)$. To understand its significance, note that the \sl electromagnetic momentum density \rm is given by\footnote{In SI units \ci[page 261]{J99}, the electromagnetic momentum density is $\3S/c^2$.
} 
$\3S\0x/c$, so $(\3S/c,\5U)$ is the energy-momentum density of the field.
For a relativistic \sl particle \rm with energy $E$ and momentum $\3p$, we have the analog of \eq{pos},
\begin{align*}
E^2-c^2\3p^2\ge 0,
\end{align*}
The mass $m$ and velocity $\3v$ of the particle are given by
\begin{align}\lab{Epv}
m=c^{-2}\sr{E^2-c^2\3p^2}\ \ \hbox{and}\ \ \3v=\frac{\3p}{E/c^2}=\frac{c^2\3p}E,
\end{align}
which implies that $\3v^2\le c^2$. Thus it makes sense to define \ci{K11} the \sl electromagnetic inertia density \rm $\5I\0x$ by
\begin{align}\lab{I}
\bx{\5I=c^{-2}\sr{\5U^2-\3S^2}=\frac1{2c^2}\sr{(\3E^2-\3B^2)^2+4(\3E\cdot\3B)^2}}
\end{align}
and the \sl electromagnetic energy flow velocity \rm $\3v\0x$ by
\begin{align}\lab{vel}
\bx{\3v=\frac{c\3S}{\5U}\imp  v\=|\3v|\le c.}
\end{align}
It will be useful also to define the \sl reactive (rest) energy density \rm 
\begin{align}\lab{Rt}
\5R\xt=c^2\5I\xt=\sr{\5U\xt^2-\3S\xt^2}.
\end{align}
Since $\3E^2-\3B^2$ and $\3E\cdot\3B$ are the two Lorentz invariants of the field, $\5I\0x$ and $\5R\0x$ are \sl local, Lorentz-invariant (scalar) spacetime fields. \rm Their physical significance, as well as that of $\3v\0x$, is the subject of this paper. Our basic theme can be summarized by the following local statement in spacetime:

\it If $|\3S\0x|< \5U\0x$, then the energy flow at $x$ does not carry away \rm all \it of the energy, leaving positive rest (reactive) energy and inertia densities at $x$. \rm

Reactive energy is an important topic in antenna theory \ci{Y96}. If an antenna generates a great deal of reactive energy, this slows down the transmission of energy and makes the antenna inefficient. Effectively, $\5R$ forms an elastic `cushion' between the source and the far-zone radiation field. While reactive energy is defined only for narrow-band fields in the literature, our definition \eq{Rt} extends this concept to arbitrary electromagnetic fields. In fact, since narrowband approximations are closely related to nonrelativistic limits \ci{K96} and $\5R$ is Lorentz-invariant, it can be said to be the ultimate `wideband' definition of reactive energy.

The quantities $\5U, \3S$ and $\5R$ can be expressed succinctly in terms of the pair of complex conjugate vector fields
\begin{align}\lab{F0}
\3F\0x=\3E\0x+i\3B\0x,\qq \3F\0x^*=\3E\0x-i\3B\0x
\end{align}
as follows:
\begin{align}\lab{ES1}
\5U=\frac{|\3F|^2}2&& \3S=\frac{\3F^*\times\3F}{2i} && \5R=\frac{|\3F^2|}{2}
\end{align}
where
\begin{align*}
|\3F|^2\=\3F^*\cdot\3F\ \ \hbox{and}\ \ \3F^2\=\3F\cdot\3F.
\end{align*}
The combinations $\3E\pm i\3B$ have been called \sl Riemann-Silberstein vectors \rm \ci{B3} and \sl Faraday vectors \rm \ci{B99}. They have been rediscovered many times and were used extensively by Bateman \ci{B15}. See also \ci{K3}, where $\3F\0x$ is continued analytically to complex spacetime,  and \ci{K4}.

By \eq{I},  
\begin{align}\lab{pos1}
\5R\0x=0\iff \3E\0x^2-\3B\0x^2=\3E\0x\cdot\3B\0x=0\iff \3F^2=0.
\end{align}
An electromagnetic field with $\3F\0x^2=0$ is said to be \sl null at $x$. \rm Nullity is a local, Lorentz-invariant property. It is the \sl field \rm counterpart of  \sl masslessness \rm in a particle. Indeed, 
\begin{align}\lab{vc}
v\0x=c\iff \5I\0x=0.
\end{align}
\sl Electromagnetic energy flows exactly at the speed of light only at events $x$ where the field is null. Elsewhere, it flows at speeds less than $c$ and has a positive inertia density. \rm

Although this simple fact should be widely known in classical electrodynamics, I've been unable to find any clear reference to it in the mainstream literature and in discussions with several knowledgeable colleagues. The sole exception, to my knowledge, is a brief note in \ci[page 6]{B15}.

However, a kind of \sl average \rm energy flow velocity per period of a time-harmonic EM field is well known in the literature under the name \sl energy transport velocity.\,\rm \footnote{I thank Professor Andrea Alu for pointing this out.
}
It is connected to the above \sl instantaneous \rm flow velocity as follows. A time-harmonic field is given by
\begin{align}\lab{EBo}
\3E\xt=\re(e^{-i\o t}\3E_\o\ox),\qq \3B\xt=\re(e^{-i\o t}\3B_\o\ox),
\end{align}
where $\3E_\o\ox$ and $\3B_\o\ox$ are complex Fourier components at a positive frequency $\o$.
Thus
\begin{align}\lab{Sot}
\3S\xt=\3S_\o\ox+\3S_\o'\xt,
\end{align}
where
\begin{align}\lab{So}
\3S_\o\ox&=\frac14(\3E_\o\times\3B_\o^*+\3E_\o^*\times\3B_\o)=\frac12\re(\3E_\o\times\3B_\o^*)
\end{align}
is the \sl average \rm of $\3S\xt$ over one period $2\p/\o$ and
\begin{align}\lab{Sot1}
\3S_\o'\xt=\frac12\re\lp e^{-2i\o t}\3E_\o\ox\times\3B_\o\ox\rp
\end{align}
oscillates at the optical frequency $2\o$. Similarly,
\begin{align}\lab{Uot}
\5U\xt=\5U_\o\ox+\5U_\o'\xt
\end{align}
where
\begin{align}\lab{Uot1}
\5U_\o\ox=\frac14\lp |\3E_\o\ox |^2+|\3B_\o\ox |^2\rp
\end{align}
is the average of $\5U$ over one period and
\begin{align}\lab{Uot2}
\5U_\o'\xt=\frac14\re \lp e^{-2i\o t}\lb\3E_\o\ox^2+\3B_\o\ox^2\rb\rp
\end{align}
oscillates at $2\o$.

The energy transport velocity is now defined as the \sl ratio of the averages: \rm
\begin{align}\lab{vo}
\3v_\o\ox\=c\,\frac{\3S_\o\ox}{\5U_\o\ox}=2c\,\frac{\re(\3E_\o\ox\times\3B_\o\ox^*)}{|\3E_\o\ox|^2+|\3B_\o\ox|^2}.
\end{align}
By comparison, our instantaneous energy flow velocity \eq{vel} is given in terms of the real fields $\3E, \3B$ by
\begin{align}\lab{vel1}
\3v\xt=c\,\frac{\3S_\o\ox+\3S_\o'\xt}{\5U_\o\ox+\5U_\o'\xt}=2c\,\frac{\3E\xt\times\3B\xt}{\3E\xt^2+\3B\xt^2}.
\end{align}
Obviously \eq{vel1} cannot  be recovered from \eq{vo}, but since vibrations at optical frequencies are generally unobservable, it might be argued that $\3v_\o\ox$ suffices for all practical purposes. 

However, it cannot be claimed that $\3v_\o\ox$ is the time average of $\3v\xt$ because \sl the ratio of averages is generally not equal to the average of ratios. \rm While $\3v_\o$ is more easily computed than the time average of $\3v$, it is not a good approximation to the latter under all circumstances.

Furthermore,  $\3v_\o$ ignores some time-dependent aspects of energy transport which help explain how the energy can flow at speeds $v<c$ even though the \sl waves \rm communicating it  propagate at $c$. This will be illustrated by studying several well-known systems.

\section{Time-harmonic electric dipole field}

The field of an oscillating electric dipole \ci{J99} with frequency $\o>0$ is given by \eq{EBo} with
\begin{align}\lab{EBdp}
\3E_\o\ox&=k^2\frac{e^{ikr}}{4\p r}\,\3e\ox, && \3e\ox=2\l(\fr z^2-i\fr z)\bh r+(1+i\fr z-\fr z^2)\3p_\perp\\
\3B_\o\ox&=k^2\frac{e^{ikr}}{4\p r}\,\3b\ox, && \3b\ox=(1+i\fr z)\bh r\times\3p,\nt
\end{align}
where $k=\o/c$ is the wavenumber, $\3p$ is a real electric dipole moment, 
\begin{align*}
\3p_\perp\=\bh r\times(\3p\times\bh r)=\3p-\l\bh r,\qq \l=\bh r\cdot\3p=p\cos\q,\qq p=|\3p|
\end{align*}
and
\begin{align*}
\fr z=\frac1{kr}
\end{align*}
is a dimensionless \sl zone parameter: \rm $\fr z\to\infty$ in the near zone, and $\fr z\to 0$ in the far zone. A straightforward computation gives
\begin{align}\lab{ebu}
&\3e\times\3b^*=(1+i\fr z^2)p^2\sin^2\q\,\bh r+2i\fr z(1+\fr z^2)p\cos\q\,\3p_\perp\\
&|\3e|^2+|\3b|^2=p^2[2\sin^2\q+4\fr z^2\cos^2\q+\fr z^4(3\cos^2\q+1)].\nt
\end{align}
Since all but the first term in $\3e\times\3b^*$ is imaginary, we get
\begin{align}\lab{Sdp}
\3S_\o\ox&=\frac{k^4p^2\sin^2\q}{32\p^2r^2}\,\bh r\ \ \hbox{where}\ \ \bh r=\frac{\3x}r\\
\5U_\o\ox&=\frac{k^4p^2}{32\p^2r^2}\lb\sin^2\q+2\fr z^2\cos^2\q+\frac12\fr z^4(3\cos^2\q+1)\rb.\nt
\end{align}
The average energy flow velocity over one period is therefore
\begin{align}\lab{vo3}
\bx{\3v_\o\ox=\frac{c\,\bh r}{1+2\fr z^2\cot^2\q+\frac12\fr z^4(3\cot^2\q+\csc^2\q)}\=v_\o(r,\q)\bh r.}
\end{align}
 
\begin{figure}[ht]
\begin{center}
\includegraphics[width=1.5 in]{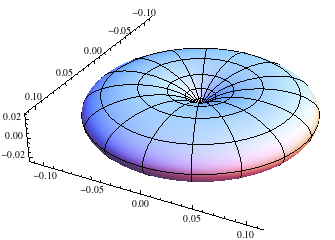}
\includegraphics[width=1.5 in]{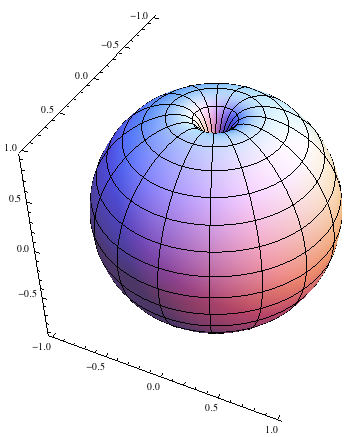}
\caption{\small The average energy flow velocity $v_\o$ per period in the near zone (left, $\fr z=2$) and the far zone  (right, $\fr z=.05$).  $v_\o$ is small and flat in the near zone, approaching $c$ in the far zone except for the poles, where it vanishes.}
\label{F:v}
\end{center}
\end{figure}

The dipole energy thus flows with an average speed $v_\o<c$, approaching $c$ only in the far zone. In the near zone it clusters near the equatorial plane $\q=\p/2$. On any sphere of constant $r$, $v_\o$ vanishes at the poles and increases monotonically towards the equator, as shown in Figure \ref{F:v}.  
 
\begin{figure}[!]
\begin{center}
\includegraphics[width=1 in]{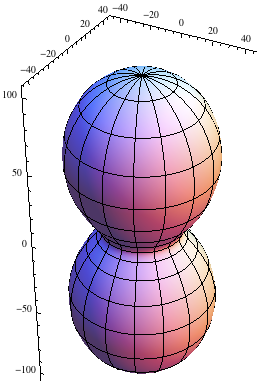}\sh2
\includegraphics[width=1 in]{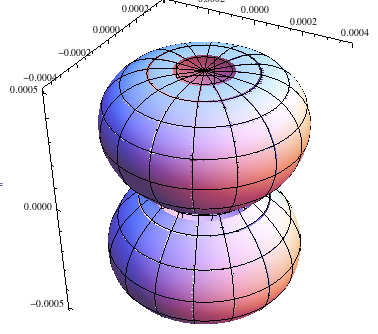}
\caption{\small The average reactive energy density $\5R_\o$ with $p=k=1$ in the near zone (left, $\fr z=5$) and the far zone  (right, $\fr z=.5$). In the near zone, it clusters around the $z$-axis  because $v_\o=0$ there, so the most energy is left behind.}
\label{F:Reactive}
\end{center}
\end{figure}
Equations \eq{Sdp} show that while $r^2\3S_\o$ is \sl zone-independent, \rm $r^2 \5U_\o$ increases monotonically as we approach the origin.  We interpret 
\begin{align}\lab{Ro}
\5R_\o\ox=c^2\5I_\o\ox\=\sr{\5U_\o\ox^2-\3S_\o\ox^2}
\end{align} 
as the \sl average reactive energy density \rm  per period. 

\sl While the radiating wave front propagates at $c$, energy is left behind wherever $|\3S_\o\ox |<\5U_\o\ox $. The abandoned energy is reactive. \rm 

Plots of $\5R_\o$ in the near and far zones are given in Figure \ref{F:Reactive}.

\section{Time-dependent electric dipole field}\label{E:dpt}

This example explains the seeming contradiction that while the time-domain dipole \sl fields \rm propagate at speed $c$, their \sl instantaneous energy density \rm generally flows at $|\3v|<c$.
Consider a general time-dependent electric dipole moment $f\0t\3p$ fixed at the origin. Its field is
\begin{align*}
\3E\xt&=\frac1{\p}\re\!\int_0^\infty\!\dd\o\,e^{-i\o t}\1f\0\o\3E_\o\ox\\
\3B\xt&=\frac1{\p}\re\!\int_0^\infty\!\dd\o\,e^{-i\o t}\1f\0\o\3B_\o\ox
\end{align*} 
where $\1f$ is the Fourier transform of $f$. Inserting \eq{EBdp}, we find
\begin{align}\lab{dptd0}
\3E\xt&=\ell(r, t_r)p\cos\q\,\bh r-e(r,t_r)\3p_\perp &&
\3B\xt=b(r, t_r)\3p\times\bh r
\end{align}
where $t_r=t-r/c$ is the retarded time and
\begin{align}\lab{leb}
\ell(r, t_r)&=\frac{f'(t_r)}{2\p cr^2}+\frac{f(t_r)}{2\p r^3}\\
b(r, t_r)&=\frac{f''(t_r)}{4\p c^2r}+\frac{f'(t_r)}{4\p cr^2}\nt\\
e(r, t_r)&=b(r, t_r)+\frac{f(t_r)}{4\p r^3}.\nt\\
\end{align}
Note that $\ell$ represents the \sl longitudinal electric component \rm  of the field while $e$ and $b$ represent its transversal electric and magnetic components. Thus
\begin{align}\lab{Ftd}
\3F=\3E+i\3B=\ell p\cos\q\,\bh r-e\3p_\perp+ib\3p_\perp\times\bh r
\end{align}
and we obtain
\begin{align}\lab{Std}
\3S&=e b p^2\sin^2\q\,\bh r+b\ell p\cos\q\,\3p_\perp\= S_r\bh r+S_\perp\3p_\perp\\
\5U&=\frac12|\3F|^2=\frac{p^2}2\lb(e^2+b^2)\sin^2\q+\ell^2\cos^2\q\rb\!. \lab{Utd}
\end{align}
$\3S$ has some remarkable properties.
\begin{itemize}
\item The second term in \eq{Std}, determined by the radial component $\ell$, is \sl transversal. \rm  It corresponds to two of the imaginary terms in \eq{ebu} which were left out when we computed the \sl average \rm energy flow per period in \eq{Sdp}. Hence \it not all of the energy is flowing outwards. \rm 

\item In the far zone we have
\begin{align*}
eb\sim\lp\frac{f''(t_r)}{4\p c^2r}\rp^2\ge 0\ \ \hbox{and}\ \ b\ell=\5O(r^{-3}),
\end{align*}
so the radial term in $\3S$ dominates and the energy is flowing \sl outwards. \rm 
\item In the near zone, $eb$ may be \sl negative, \rm hence energy may flow \sl inwards \rm as well as transversally. This will be confirmed below.
\end{itemize}

The expression for the instantaneous speed $v=c|\3S|/\5U$ is rather complicated. However, the reactive energy density
\begin{align}\lab{Rt1}
\5R=\frac12|\3F^2|=\frac{p^2}2\!\bigm|\!\!(e^2-b^2)\sin^2\q+\ell^2\cos^2\q\!\bigm|\!
\end{align}
gives a simple expression for $v$ in the relativistic form
\begin{align}\lab{vtd}
\bx{\sr{1-\frac{v^2}{c^2}}=\frac{\5R}{\5U}
=\frac{\!\bigm|\!\!(e^2-b^2)\sin^2\q+\ell^2\cos^2\q\!\bigm|\!\!}{(e^2+b^2)\sin^2\q+\ell^2\cos^2\q}\,.}
\end{align}
Recall  that for the time-harmonic dipole we had $\3v_\o=\30$ on the dipole axis. This extends to the \sl instantaneous \rm velocity for the pulsed dipole field:
\begin{align*}
\sin\q=0\imp \sr{1-\frac{v^2}{c^2}}=1\imp \3v=0
\end{align*}
provided $\ell\ne0$, which can occur only at special values of $r$ and $t$ as shown below.
Since the expression on the right of \eq{vtd} is nonnegative, we have $v\le c$ as required. 
Furthermore, 
\begin{align*}
v=c\iff (e^2-b^2)\sin^2\q+\ell^2\cos^2\q=0.
\end{align*}
For a generic pulse $f$, this can only be satisfied on \rm sets of zero measure \rm in spacetime. For example, 
\begin{align}\lab{ebl0}
v=c\ \ \hbox{if}\ \  e^2-b^2=\ell=0,
\end{align}
which looks a little like a scalar version of the nullity condition \eq{pos1}. Such conditions can be satisfied only for special values of $r,\q$ and $t$. For example, if $\ell(r,t)=0$ for a \sl finite \rm time interval, then
\begin{align*}
f'\0t=-\frac cr f\0t\imp f\0t=A e^{-ct/r}.
\end{align*}
This is unacceptable as a pulse function because it depends on $r$.\footnote{It also grows exponentially for $t<0$, but we could set $f=0$ for $t<0$.
}
This shows that $\ell$ can have only \sl isolated \rm zeros.
Similarly, $e^2-b^2$ can have only isolated zeros. Thus we have
\begin{align}\lab{vlc}
v\xt<c\ \ \hbox{almost everywhere (a.e.) in spacetime.}
\end{align}
\sl The energy of the pulsed dipole field flows at speeds less than $c$ a.e. \rm

This statement feels uncomfortable because \eq{dptd0} and \eq{leb} show that the fields $\3E,\3B$, hence also $\5U$ and $\5R$, depend on $t$ only through the retarded time $t-r/c$. If $f$ is a sharp pulse, then so are $\ell, e, b$. It follows that $\5U$ and $\3S$, like the fields, are small unless $r\app ct$. How can this be reconciled with \eq{vlc}? To investigate this, consider the modulated Gaussian pulse
\begin{align}\lab{modgauss}
f\0t=g\0t\cos\o t\ \ \hbox{where}\ \ g\0t=e^{-\k t^2/2},\ \k>0,\ \o>0
\end{align}
and set $c=1$ and $p=1$ for notational convenience.
Then
\begin{align*}
f'\0t&=-\k t g\0t\cos\o t-\o g\0t\sin\o t\\
f''\0t&=(\k^2t^2-\k-\o^2)g\0t\cos\o t+2\k\o t g\0t\sin\o t,
\end{align*}
hence
\begin{align}\lab{mus}
4\p r^3 \ell(r,t)&=\m_\ell(r,t) g\0t\\
4\p r^3 b(r,t)&=\m_b(r,t) g\0t\nt\\
4\p r^3 e(r,t)&=\m_e(r,t) g\0t\nt
\end{align}
with
\begin{align}\lab{mus1}
\m_\ell(r,t)&=2(1-\k rt)\cos\o t-2\o r\sin\o t\\
\m_b(r,t)&=[\k^2r^2t^2-(\k+\o^2) r^2-\k r t]\cos\o t+[2\k\o r^2t-\o r]\sin\o t\nt\\
\m_e(r,t)&=\m_b(r,t)+\cos\o t.\nt
\end{align}

\begin{align}\lab{SUR}
\3S\xt&=\lb\frac{g(t_r)}{4\p r^3}\rb^2\3S_o(\3x, t_r)\\
\5U\xt&=\lb\frac{g(t_r)}{4\p r^3}\rb^2\5U_o(\3x, t_r)\nt\\
\5R\xt&=\lb\frac{g(t_r)}{4\p r^3}\rb^2\5R_o(\3x, t_r)\nt
\end{align}
where
\begin{align}\lab{SRUo}
\3S_o(\3x, t_r)&=\m_e\m_b\sin^2\q\,\bh r+\m_b\m_\ell\cos^2\q\,\3p_\perp\\
\5U_o(\3x, t_r)&=\frac12\lb\lp\m_e^2+\m_b^2\rp\sin^2\q+\m_\ell^2\cos^2\q\rb\nt\\
\5R_o(\3x, t_r)&=\frac12\!\bigm|\!\lp\m_e^2-\m_b^2\rp\sin^2\q+\m_\ell^2\cos^2\q\!\bigm|,\nt
\end{align}
with all expressions on the right evaluated at the retarded time $t_r$. While $\3S, \5U$ and $\5R$ all contain the Gaussian factor $g(t_r)^2=e^{-\k t_r^2}$, this factor  \sl cancels \rm in the ratios
\begin{align}\lab{vtd2}
\3v=2c\,\frac{\m_e\m_b\sin^2\q\,\bh r+\m_b\m_\ell\cos^2\q\,\3p_\perp}
{\lp\m_e^2+\m_b^2\rp\sin^2\q+\m_\ell^2\cos^2\q}\=v_r\bh r+v_\perp\3p_\perp
\end{align}
and
\begin{align}\lab{vtd3}
\sr{1-\frac{v^2}{c^2}}=\frac{\5R_o}{\5U_o}
=\frac{\!\!\bigm|\!\!\lp\m_e^2-\m_b^2\rp\sin^2\q+\m_\ell^2\cos^2\q\!\!\bigm|\!\!}
{\lp\m_e^2+\m_b^2\rp\sin^2\q+\m_\ell^2\cos^2\q}\,.
\end{align}
These ratios depends only on the relative sizes of the factors $\m_e, \m_b$ and $\m_\ell$ at time $t_r$. \it The pulse $g(t_r)^2$ cancels in \eq{vtd3}. \rm 

It can be argued that for \sl large \rm $|t_r|$, both $\3S$ and $\5U$ are extremely small and hence their ratio has little meaning. When almost no energy is flowing, the speed of its flow is largely of academic interest. This is certainly true in the general case \eq{vtd} at times when $\5R$ and $\5U$ vanish identically.

Equation \eq{vtd3} confirms that $v$ vanishes on the dipole axis ($\sin\q=0$). It also vanishes for special values of $r,\q$ and $t$, for example when $\m_b=0$ or when $\m_\ell=\m_e=0$. 
There, the instantaneous energy is entirely reactive. 

\begin{figure}[ht]
\begin{center}
\includegraphics[width=2 in]{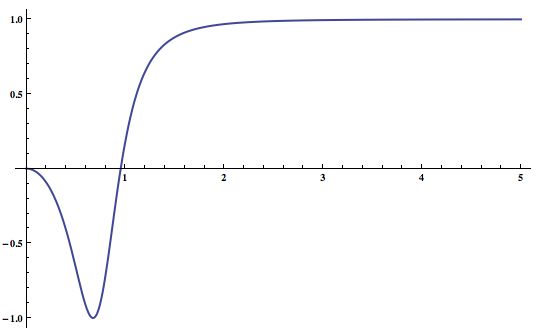}\sh2
\includegraphics[width=2 in]{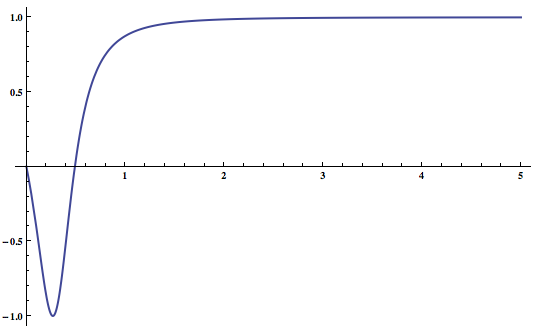}
\caption{\small Plots of the radial energy flow velocity $v_r$ for the Gaussian pulse \eq{modgauss} as a function of $r$ with $t_r=0$ (left) and $t_r=1$ (right). We have set $\q=\p/2$, $\o=1$ and $\k=0.1$. When $v_r<0$, the energy flows towards the origin.}
\label{F:Fig_vr}
\end{center}
\end{figure}

\bf Remark 1. \rm
In the discussion below Equation \eq{Std} we have noted that when $eb<0$, the field energy flows \sl inwards. \rm  For the modulated Gaussian pulse \eq{modgauss}, this occurs when 
\begin{gather*}
\m_e(r, t_r)\m_b(r, t_r)=\m_b(r, t_r)^2+\m_b(r, t_r)\cos\o t_r<0.
\end{gather*}
Since $\m_b(r, t_r)=\5O(r^2)$ as $r\to\8$ at given $t_r$, it follows that
\begin{align*}
r\to\8\imp\m_e(r, t_r)\m_b(r, t_r)\sim\m_b(r, t_r)^2\ge 0\imp v_r\ge 0.
\end{align*}
But in the near zone, the radial velocity $v_r$ in \eq{vtd2} can be negative, in which case the energy  flows inwards. Figure \ref{F:Fig_vr} shows the behavior of $v_r$ as a function of $r$ on the equatorial plane $\q=\p/2$. Note that the energy flows inwards for small values of $r$ and then quickly transitions to flowing outward at $v_r\app c$. The ingoing flow does not show up in the time-averaged velocity \eq{vo3}, which is outgoing in all zones. 

\bf Remark 2. \rm
For a general pulse function $f\0t$, no simple relations like \eq{mus} exist. Nevertheless, the velocity $\3v=c\3S/\5U$ depends on the relative sizes of $f'', f'$ and $f$ through the ratio \eq{vtd}, and not merely on $f\0t$. While the pulse does not actually cancel, as it did for the Gaussian, the numerator and denominator of \eq{vtd} are \sl both \rm small when $t_r$ is sufficiently large, hence nothing definite can be said about $\3v$ without knowing the relative sizes of $\ell, e, b$. This explains how the energy can flow at speeds less than $c$ while the fields propagate at $c$.

\it However,  the value of $\3v\xt$ says nothing about the \rm quantity \it of energy flowing at this velocity. \rm For example, we find that 
\begin{align*}
\sin\q\ne 0\ \ \hbox{and}\ \ kt_r^2\gg 1\imp v\to c.
\end{align*}
Although almost no energy remains when $t_r$ is large, what little there is flows nearly at the speed of light provided we are not on the dipole axis. As an extreme case of this, let $f\0t$ vanish outside the interval $T=[t_1, t_2]$. Then $\3v\xt$ is undefined for $t\notin T$.
 
\begin{figure}[ht]
\begin{center}
\includegraphics[width=.7 in]{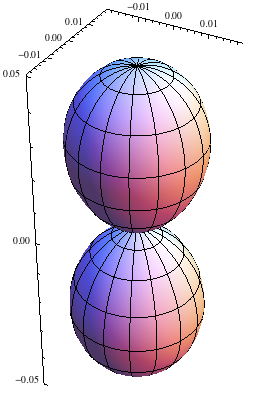}
\includegraphics[width=.7 in]{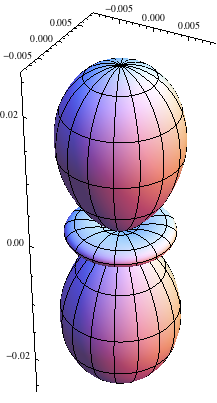}
\includegraphics[width=1.2 in]{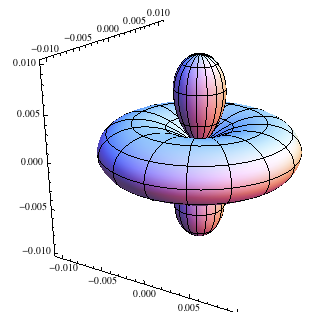}
\includegraphics[width=1.1 in]{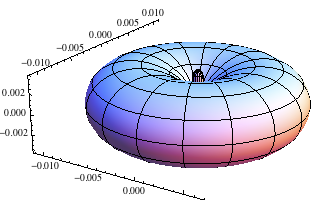}
\includegraphics[width=.8 in]{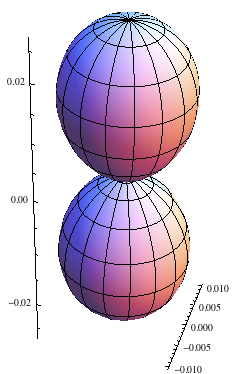}
\caption{\small \sl Left to right: \rm The reactive energy density $\5R$ of the modulated Gaussian dipole field with $r=1$, $\o=1, \k=0.1$ at $t=0.5, 1,1.3, 1.5, 3$. The vertical lobes and horizontal tubes are due to the $\sin^2\q$ and $\cos^2\q$ terms in \eq{SRUo}, which take turns dominating because $\m_\ell, \m_b$ and $\m_e$ all oscillate with period $\p$. 
}
\label{F:Rdpt}
\end{center}
\end{figure} 
\begin{figure}[ht]
\begin{center}
\includegraphics[width=1 in]{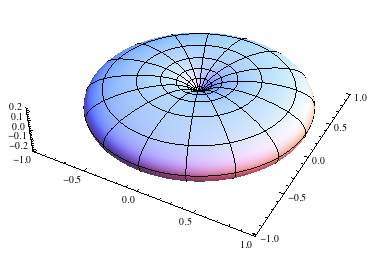}\sh{-2}
\includegraphics[width=1 in]{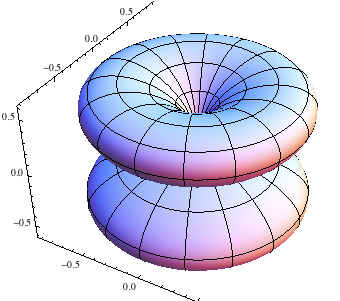}\sh{-1}
\includegraphics[width=1 in]{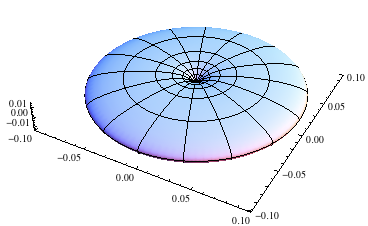}\sh{-2}
\includegraphics[width=.9 in]{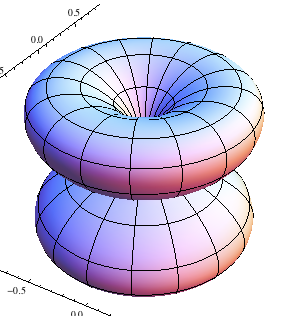}\sh{-1}
\includegraphics[width=1 in]{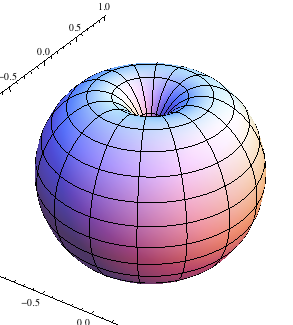}
\caption{\small \sl Left to right: \rm The energy flow speed $v$ in \eq{vtd3} of the modulated Gaussian dipole field with $r=1$, $\o=1, \k=0.1$ at $t=0.5, 1,3,4, 30$. Note that $v$ has a zero near $t=3$, and $v\to c$ as $t\to\8$ even though $\5R,\5U\to0$.}
\label{F:vdpt}
\end{center}
\end{figure}

Figures \ref{F:vdpt} and \ref{F:Rdpt} show the reactive energy density energy $\5R$ and the energy flow speed
\begin{align}\lab{vtd4}
v\xt=c\sr{1-\5R^2/\5U^2}
\end{align} 
of the modulated Gaussian dipole field on the unit sphere $r=1$ at various times. As $t\to\8$, $v\to c$ outside the dipole axis even though the \sl quantity \rm of energy flowing vanishes.

\section{Standing plane wave}\label{E:standing}

Our final example demonstrates an important feature of null fields: \sl they do not interfere with themselves as they propagate. \rm 

The simplest examples of null fields are \sl plane waves \rm with real wave vectors.\footnote{In the \sl plane-wave spectrum representation \rm \ci{HY99} (also called the \sl angular spectrum representation\rm), electromagnetic fields are expressed as superpositions of plane waves with \sl complex \rm wave vectors. Such `inhomogeneous' plane waves are not null since they include non-propagating \sl evanescent waves. \rm
} 
Consider a pair of  linearly polarized plane waves propagating along the positive and negative $z$-axis,
\begin{align*}
\3E_\pm\0x=\bh x E\cos(ct\mp z),\qq \3B_\pm=\pm\bh y E\cos(ct\mp z).
\end{align*}
These are solutions of Maxwell's equations with complex representations
\begin{align}\lab{Fpm}
\3F_\pm\=\3E_\pm+i\3B_\pm=(\bh x\pm i\bh y)E\cos(ct\mp z).
\end{align}
The energy-momentum of $\3F_\pm$ is
\begin{align*}
\5U_\pm=E^2\cos^2(ct\mp z) &&
\3S_\pm=\pm\bh z E^2\cos^2(ct\mp z),
\end{align*}
hence the inertia density and energy flow velocity are
\begin{align*}
\5I_\pm\=\frac1{2c^2}|\3F_\pm^2|=0,\qq \3v_\pm\=\frac{c\3S_\pm}{\5U_\pm}=\pm c\,\bh z.
\end{align*}
Thus $\3F_\pm$ are \sl null fields \rm moving in the $\pm z$ direction at speed $c$. 

\begin{figure}[!]
\begin{center}
\includegraphics[width=2 in]{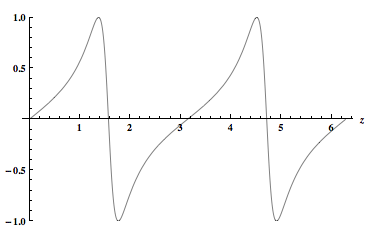}
\includegraphics[width=2 in]{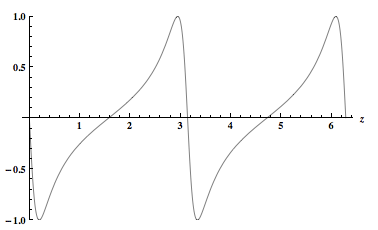}
\caption{\small Plots of $v(z,t)$ at $t_1=\p/16$ (\sl left\rm) and half a cycle later, at $t_2=9\p/16$ (\sl right\rm). On the left, the energy travels to the right in the intervals $(0,\p/2)$ and $(\p, 3\p/4)$, and to the left in $(\p/2,\p)$ and $(3\p/4, 2\p)$. On the right, it travels in the opposite directions. Furthermore, $v(z,0)=v(z,\p)=0$ and $v(z,t)$ jumps discontinuously across the odd nodes $z_{2n+1}$ as $t\to 0$ and across the even nodes $z_{2n}$ as $t\to\p$. Note that $v$ is not at even close to harmonic.}
\label{F:Fig_t}
\end{center}
\end{figure}

Now consider the \sl standing wave \rm
\begin{align}\lab{F2}
\3F&\=\3F_++\3F_-=(\bh x+ i\bh y)E\cos(ct-z)+(\bh x-i\bh y)E\cos(ct+z),
\end{align}
which can also be written as
\begin{align*}
\3F&=\3E+i\3B\ \ \hbox{where}\ \  \3E=2\bh x E\cos ct\cos z,\  \3B=2\bh yE\sin ct\sin z.
\end{align*}
Note that \eq{F2} is expressed in terms of the traveling waves $\cos(ct\pm z)$ while $\3E$ and $\3B$ are not. It will therefore be more revealing to use $\3F$. We find
\begin{align*}
\5I(z,t)&=\frac{|\3F^2|}{2c^2}=2(E^2/c^2)|\cos(ct-z)\cos(ct+z)|\\
\5U(z,t)&=\frac{\3F^*\cdot\3F}2=E^2\lb\cos^2(ct-z)+\cos^2(ct+z)\rb\\
\3S(z,t)&=\frac{\3F^*\times\3F}{2i}=\bh zE^2\lb\cos^2(ct-z)-\cos^2(ct+z)\rb,
\end{align*}
and the energy flow velocity is
\begin{align}\lab{vzt}
\bx{\3v(z,t)\=\frac{c\3S}{\5U}=\bh z v(z,t),\ 
v(z,t)=c\,\frac{\cos^2(ct-z)-\cos^2(ct+z)}{\cos^2(ct-z)+\cos^2(ct+z)}.}
\end{align}
As expected, $|\3v|\le c$. Note that $v$ is periodic of period $\p/c$, and
\begin{align*}
\3v=\30\iff \cos^2(ct-z)=\cos^2(ct+z)\iff ct+z=\pm(ct-z)+n\p.
\end{align*}
Hence $\3v$ has \sl fixed nodes in space and time, \rm as seen in Figure \ref{F:Fig_t}:
\begin{align}\lab{nodes}
\3v(z,t)=\30\iff z=\frac{n\p}2\=z_n\ \ \hbox{or}\ \ \ ct=\frac{n\p}2\= ct_n,\qq  n\in\4Z.
\end{align}
Since $v(z,t)$ changes sign at $z_n$, the energy is \sl reflected \rm at the nodes. 

\it The energy oscillates back and forth between successive nodal planes, and $v(z,t)$ oscillates between $\pm c$ at any $z$ which is not a node. \rm

\begin{figure}[!]
\begin{center}
\includegraphics[width=3 in]{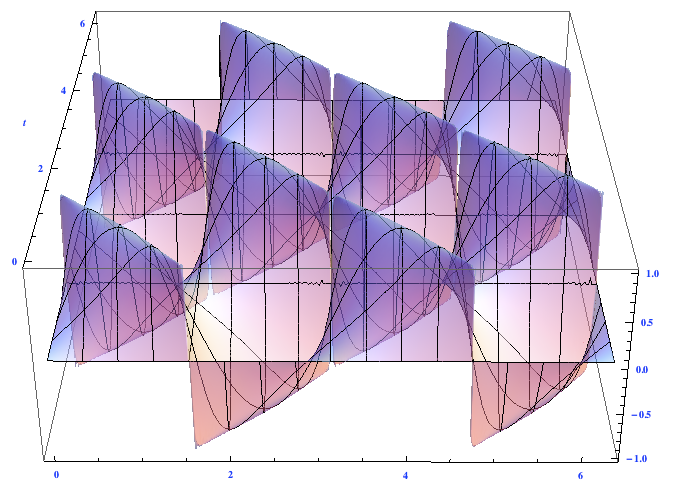}
\caption{\small Plot of $v(z,t)$ \eq{vzt} showing the nodes \eq{nodes} and confirming \eq{vz}.}
\label{F:Fig_Standing}
\end{center}
\end{figure}
For non-nodal $z$, we have
\begin{align}\lab{vz}
\cos^2(ct+z)=0\iff z\0t=(n+\tfrac12)\p-ct&\iff v=c\\
\cos^2(ct-z)=0\iff z\0t=(n+\tfrac12)\p+ct&\iff v=-c.\nt
\end{align}
That is, if we begin at the node $z\00=z_{2n+1}=(n+\tfrac12)\p$ and travel to the left at speed $c$, we see the velocity of $\3F_+$ except when crossing the nodes. If instead we travel to the right at speed $c$, we see the velocity of $\3F_-$. This behavior can be seen in Figures \ref{F:Fig_t} and \ref{F:Fig_Standing}. Note that $v(z,t)$ jumps \sl discontinuously \rm between $\pm c$ across the nodes. This is because rather than crossing the nodes, the velocity builds up to $\pm c$ and is then immediately reflected to $\mp c$.

Had we used the time-averaged formula, we would have obtained the expected result $\3v_\o=\30$. However, \eq{vel} gives a detailed, instantaneous picture of the movement of energy.

\section{The incoherence of electromagnetic fields}

The above examples show that we must distinguish between the propagation speed of the \sl fields \rm $(\3E, \3B)$ and the flow velocity of their \sl energy. \rm  Let us think about this from a purely \sl mathematical \rm point of view.

\bul The fields satisfy homogeneous wave equations outside their sources, hence they propagates at speed $c$.

\bul On the other hand, $(\3S,\5U)$ are quadratic functions of the fields and satisfy Poynting's conservation law outside of sources,
\begin{align}\lab{conserv}
\pl_t \,\5U+c\div\3S=\pl_t \,\5U+\div(\3v\, \5U)=0.
\end{align}
Hence $\5U$ behaves like the density of a \sl compressible fluid \rm flowing with velocity $\3v$. Although \eq{conserv} follows from Maxwell's equations, there is no \sl a priori \rm reason why the two speeds should be equal. Indeed, as we have shown, they are in general different; only in the far zone\footnote{Provided, of course, that a far zone exists. When the sources have an infinite extent or are `at infinity' as in the case of plane waves, no `far zone' may exist. For example, the standing wave \eq{F2} has no far zone, and the traveling plane waves \eq{Fpm} have \sl all \rm of spacetime as their far zone.
}
does $v\to c$.

A rough way to understand why $v<c$ is by analogy with water waves. The \sl mass \rm carried by the waves has a definite speed at each point and time, but this need not coincide the propagation speed of the wavefronts. 

\it But null fields are an exception: all of the energy is carried away by the wave since $v=c$. For them, the energy flows in unison with the waves, hence they play a very special role in electrodynamics.\rm

The following definition is motivated by the fact that the electromagnetic energy propagates \sl coherently \rm with the field if and only if $\3F\0x^2=0$. In fact, the traveling plane waves \eq{Fpm} are \sl mutually coherent \rm in this sense even at different spacetime points:
\begin{align*}
\3F_\pm\0x=(\bh x\pm i\bh y)E\cos(ct\mp z)\imp \3F_\pm\0x\cdot\3F_\pm(y)=0\ \forall x, y\in\rr4
\end{align*}
since $(\bh x\pm i\bh y)^2=0$.

\begin{defin}\label{D:}\rm
The \it incoherence \rm of an electromagnetic field $\3F=\3E+i\3B$ is the complex function
\begin{align}\lab{Incoh}
\fr I(x,y)=\3F\0x\cdot\3F(y),
\end{align}
which is expressed in terms of the fields $\3E, \3B$ as
\begin{align*}
\fr I(x,y)=\3E\0x\cdot\3E\0y-\3B\0x\cdot\3B\0y+i\3E\0x\cdot\3B\0y+i\3B\0x\cdot\3E\0y.
\end{align*}
\end{defin}

$\fr I$ measures the incoherence of $\3F$ across space and time by comparing $\3F$ at two different events. It can be related to the \sl coherence functions \rm of statistical optics   \ci{W7} as follows. 

If $\3F$ is time-harmonic,
\begin{align*}
\3E\xt=2\re (e^{-i\o t}\3e\ox),\qq \3B\xt=2\re(e^{-i\o t}\3b\ox),
\end{align*}
then the \sl average \rm $\fr I_\o$ of the \sl equal-time \rm incoherence function over one period is given by
\begin{align}\lab{Incoh0}
\fr I_\o(\3x,\3y)
&=\3e\ox\cdot\3e(\3y)^*+\3e\ox^*\cdot\3e(\3y)-\3b\ox\cdot\3b(\3y)^*-\3b\ox^*\cdot\3b(\3y)\\
\qq& +i\3e\ox\cdot\3b(\3y)^*+i\3e\ox^*\cdot\3b(\3y)+i\3b\ox\cdot\3e(\3y)^*+i\3b\ox^*\cdot\3e(\3y).\nt
\end{align}
This can be expressed in the compact form
\begin{align}\lab{incoh1}
\fr I_\o(\3x,\3y)=\3f_+\ox\cdot\3f_-(\3y)^*+\3f_-\ox^*\cdot\3f_+(\3y),
\end{align}
where
\begin{align*}
\3f_+\ox=\3e\ox+ i\3b\ox\ \ \hbox{and}\ \ \3f_-\ox=\3e\ox-i\3b\ox
\end{align*}
are independent because $\3e$ and $\3b$ are complex. 

The connection with the coherence functions of statistical optics is obtained by extending $\fr I$ to \sl random fields, \rm where we have an ensemble of fields $\3F\0x$ and its incoherence is defined by the ensemble average
\begin{align}\lab{incoh2}
\fr I(x,y)&=\la\3F\0x\cdot\3F\0y\ra.
\end{align}
Equation \eq{incoh1} shows that for a random time-harmonic field, $\la\fr I\ra_\o$ is a specific combination of coherence functions for the electric and magnetic fields. But while coherence functions are designed to test the \sl correlation \rm of a set of fields, our incoherence function represents their \sl electric-magnetic imbalance, \rm in the sense that a random field $\3F$ is \sl perfectly balanced \rm at $x$ when 
\begin{align*}
\la\3F\0x^2\ra=0,\ \ \hbox{\ie}\ \ \la\3E\0x^2\ra=\la\3B\0x^2\ra\ \ \hbox{and}\ \ \la\3E\0x\cdot\3B\0x\ra=0.
\end{align*}
This notion of balance is required for the field's energy to propagate coherently with the field. In the time-harmonic case, this means we have the following identities between the correlation functions:
\begin{align*}
\re\la\3e\ox\cdot\3e(\3y)^*\ra&=\re\la\3b\ox\cdot\3b(\3y)^*\ra\\
\re\la\3e\ox\cdot\3b(\3y)^*\ra&=-\re\la\3b\ox\cdot\3e(\3y)^*\ra.
\end{align*}
Whereas the coherence functions of statistical optics are defined only for random fields, we have seen that the incoherence function has a deep significance even for a deterministic field. 

A generic electromagnetic field in free space is null along a set of 2-dimensional hypersurfaces $\5S$ in spacetime since $\3F\xt^2=0$ imposes two real conditions on the four spacetime 
variables $\xt$.\footnote{We have seen an example of this in Section \ref{E:dpt}, where the coherence condition \eq{ebl0} for the time-dependent electric dipole field can only be satisfied on isolated surfaces.
}
The time slices $\5S_t$ of $\5S$ are (generically) curves in space, and these curves evolve with 
$t$.\footnote{Bialynicki-Birula \ci{B3} calls these moving curves \sl electromagnetic vortices, \rm but the `rotations' around these curves are \sl duality rotations \rm $\3F\to e^{i\f}\3F$ rather than rotations in physical space.
}
Thus, when a field is not null in an extended region of spacetime, its energy flows at the speed of light only along such curves.  Elsewhere it flows at speeds $v<c$, although $v\to c$ in the far zone.  

So far, our only example of a null field has been the traveling plane waves $\3F_\pm$ \eq{Fpm}. It is  easy to make a plane wave null because its energy flow velocity is a constant vector. 
Various other globally null electromagnetic fields are known where $\3F^2$ vanishes \sl almost everywhere, \rm with the possible exception of singularities whose supports have zero measure. Such fields  play an important role in the solution of the Einstein-Maxwell equations of general relativity \ci{B15,R61,T62, RT64}. However, no \sl extended, compactly supported sources \rm appear to be known which radiate null fields everywhere outside the source region. Such null field, called \sl coherent electromagnetic wavelets, \rm was recently constructed \ci{K11}. It is radiated by a relativistically spinning charged disk, and the radiation follows a \sl twisting null congruence \rm of light rays. This is a space-filling set of null lines (world lines of particles traveling at speed $c$) which are determined by the energy flow velocity vector $\3v=c\3S/\5U$ with $|\3v|=c$.. The existence of such null congruences is the key to the construction of nontrivial null fields.

\section*{Acknowledgements}
I thank Drs. Richard Albanese, Andrea Alu, Richard Matzner, Arje Nachman, and Arthur Yaghjian for posing challenging questions and engaging me in extended and helpful discussions on the subject of energy flow velocity. This work was supported by AFOSR Grant \#FA9550-08-1-0144.

\end{document}

%% file: B3macros.tex
\def\div{\nabla\cdot } 
\def\pl{\partial}

\def\={\equiv}

\newcommand{\xt}{(\3x,t)}

\newcommand{\ox}{(\3x)}

\newtheorem{defin}{Definition}

\newcommand{\bib}{\bibitem}

\newcommand{\nt}{\notag}
\newcommand{\ci}{\cite}
\newcommand{\lab}{\label}

\newcommand{\eq}{\eqref}

\newcommand{\bx}[1]{\boxed{\ #1\ }}

\newcommand{\lp}{\left(}
\newcommand{\rp}{ \right)}
\newcommand{\lb}{ \left[}
\newcommand{\rb}{\right]}
\newcommand{\la}{\langle\,}

\newcommand{\ra}{\,\rangle}

\newcommand{\harr}[1]{\smash{\mathop{\hbox to .5in{\ \rightarrowfill\ }}
      \limits^{#1}}}

\newcommand{\0}[1]{{(#1)}}

\newcommand{\3}[1]{{\boldsymbol #1}}

\newcommand{\bh}[1]{{\boldsymbol{\hat #1}}}

\newcommand{\fr}[1]{{\mathfrak #1}}

\newcommand{\8}{\infty}

\def\e{\varepsilon} 

\def\f{\phi}

\def\k{\kappa}

\def\l{\lambda} 
\def\m{\mu} 

\def\o{\omega} 
\def\p{\pi} 

\def\q{\theta}

\newcommand{\bul}{$\bullet\ $}

\newcommand{\db}{{\,{\rm d}\kern-1.6ex-}}
\newcommand{\dir}{{\pl\kern-1.2ex {/}}}
\newcommand{\dd}{{\rm d}}

\newcommand{\app}{\approx}

\newcommand{\ie}{{\it i.e., }}

\def\iff{\ \Leftrightarrow\ }
  
\newcommand{\imp}{\ \Rightarrow\ }

\newcommand{\lra}{\leftrightarrow}

\newcommand{\plra}{\pl^{\kern-1.25ex^\lra}}
\newcommand{\qq}{\quad} 
 
\newcommand{\re}{{\,\rm Re}\  }   

\newcommand{\rr}[1]{{{\mathbb R}^{#1}}}

\newcommand{\sh}[1]{\hskip#1ex} 
\newcommand{\sr}{\sqrt}

\def\XXint#1#2#3{{\setbox0=\hbox{$#1{#2#3}{\int}$}
     \vcenter{\hbox{$#2#3$}}\kern-.5\wd0}}

\def\VE{\vfill\eject}